# Interfacial reaction boosts thermal conductance of room-temperature integrated semiconductor interfaces stable up to 1100 °C

Zhe Cheng[1,2,5,*,‡], Xiaoyang Ji,[2,*] Zifeng Huang,[1] Yutaka Ohno,[3] Koji Inoue,[3] Yasusyohi Nagai,[3] Yoshiki Sakaida,[4] Hiroki Uratani,[4] Naoteru Shigekawa,[6] Jianbo Liang[6,‡]

[1] School of Integrated Circuits and Beijing Advanced Innovation Center for Integrated Circuits, Peking University, Beijing 100871, China

[2] Department of Materials Science and Engineering and Materials Research Laboratory, University of Illinois at Urbana-Champaign, Urbana, Illinois, 61801, USA

[3] Institute for Materials Research (IMR), Tohoku University, Ibaraki 311-1313, Japan

[4] SiC Division, Air Water Inc. 2290-1 Takibe, Toyoshina Azumino, Nagano 399-8204, Japan

[5] Frontiers Science Center for Nano-optoelectronics, Peking University, Beijing 100871, China

[6] Department of Physical Electronics, Osaka Metropolitan University, Osaka 599 – 8531, Japan

[*]These authors contribute equally.

[‡] Corresponding authors: zhe.cheng@pku.edu.cn; g21907a@omu.ac.jp

**Abstract**

Overheating has emerged as a primary challenge constraining the reliability and performance of next-generation high-performance electronics, such as chiplets and (ultra)wide bandgap electronics. Advanced heterogeneous integration not only constitutes a pivotal technique for fabricating these electronics but also offers potential solutions for thermal management. This study presents the integration of high thermal conductivity semiconductors—specifically, 3C-SiC thin films and diamond substrates—through a room-temperature surface-activated bonding technique. Notably, the thermal conductivity of the 3C-SiC films is among the highest for all semiconductor films which can be integrated near room temperature with similar thicknesses. Furthermore, following annealing, the interfaces between 3C-SiC and diamond demonstrate a remarkable enhancement in thermal boundary conductance (TBC), reaching up to approximately 300%, surpassing all other grown and bonded heterointerfaces. This enhancement is attributed to interfacial reactions, specifically the transformation of amorphous silicon into SiC upon interaction with diamond, which is further corroborated by picosecond ultrasonics measurements. Subsequent to annealing at 1100 °C, the achieved TBC (150 MW/$m^2$-K) is record-high among all bonded diamond interfaces. Additionally, the visualization of large-area TBC, facilitated by femtosecond laser-based time-domain thermoreflectance measurements, shows the uniformity of the interfaces which are capable of withstanding temperatures as high as 1100 °C. Our research marks a significant advancement in the realm of thermally conductive heterogeneous integration, which is promising for enhanced cooling of next-generation electronics.

# 1. Introduction

The rapid evolution of artificial intelligence and the internet of things has fueled a growing demand for high-performance chips. However, as Moore's law nears its limits, the cost of traditional transistor two-dimensional scaling escalates sharply, with diminishing returns in performance.[1–4] This challenge has prompted a shift towards three-dimensional stacking of multiple chips with distinct functionalities, a technique known as chiplets, which has garnered significant attention.[5–9] Nevertheless, effective thermal management within chiplets poses a formidable obstacle due to the constrained paths for heat dissipation.[10–12] One potential solution lies in the thermally conductive integration of high thermal conductivity semiconductors, offering a means to mitigate overheating issues and diminish thermal resistances within chiplets.[13] For instance, materials like diamond or SiC could supplant conventional materials such as silicon, glass, or polymers in through-silicon vias (TSV), thereby enhancing thermal conductivity and dissipating heat more efficiently.[14]

Moreover, propelled by the rapid expansion of applications such as data centers, electric vehicles, 5G stations, radar systems, and satellite communications, there is an increasing demand for size-compact and energy-efficient power and radio-frequency (RF) devices.[15,16] (Ultra)wide bandgap (WBG or UWBG) semiconductors, including GaN, SiC, and $\beta$-$Ga_2O_3$, hold particular promise for these devices owing to their high critical electric field and low specific on-resistance.[17–21] However, localized Joule heating in the channel region creates hot spots that degrade device performance and reliability.[22–24] To reduce peak temperature, devices are heterogeneously integrated with high thermal conductivity substrates such as diamond. Nevertheless, the thermal boundary conductance (TBC) of the semiconductor interfaces remains crucial for efficient heat dissipation.[22,25]

Advanced heterogeneous integration techniques, such as plasma bonding, hybrid bonding, surface activated bonding (SAB), and hydrophilic bonding, play a fundamental role in both the fabrication and thermal management of next-generation high-performance electronic devices. Previous studies have reported the integration of GaN devices and BAs substrates with an $Al_2O_3$ interfacial layer via plasma bonding.[26] GaN has also been bonded with SiC using SAB, achieving high TBC for the as-bonded interface, with further enhancement observed after annealing by reducing the amorphous interfacial layer.[27–30] Additionally, bonding of GaN with diamond has been demonstrated, albeit limited to small areas.[31–35] For instance, GaN has been bonded with single crystal diamond via SAB process using ultrathin Si interlayers, achieving TBCs exceeding 90 MW/m$^2$-K.[28] Furthermore, bonding of GaN with diamond using metallic interlayers has also been successful, resulting in TBCs exceeding 100 MW/m$^2$-K.[36] The challenges associated with direct integration of large-area GaN with diamond have led to investigations into GaN/SiC/diamond structures. Previous research has shown successful bonding of SiC and diamond with a Ti interlayer via SAB, enabling the fabrication of AlGaN/GaN HEMTs on SiC/diamond substrates, albeit with a reported TBC of only 15 MW/m$^2$-K.[37] More recently, the integration of GaN HEMTs on SiC/diamond substrates through SAB has resulted in devices with higher drain current and lower thermal resistance compared to those on SiC substrates.[34]

To achieve excellent thermal management of chiplets and power/RF electronics, three critical factors govern the heterogeneous integration of semiconductors. Firstly, the semiconductors used for integration must possess high thermal conductivity. Secondly, the integrated interfaces should exhibit high TBC, crucial for efficient heat spreading or dissipation. Thirdly, the heterostructures or composite wafers should withstand high temperatures, facilitating further device growth or fabrication.

In this study, we employ the room-temperature surface activated bonding (SAB) technique to integrate high thermal conductivity 3C-SiC thin films with monocrystalline/polycrystalline diamond substrates. The bonded 3C-SiC/diamond interfaces are investigated using ultrafast laser-based time-domain thermoreflectance (TDTR) and picosecond ultrasonic techniques to understand the effects of high-temperature annealing on thermal properties and interfacial structures. Furthermore, atomic scale characterizations are employed to investigate the structure and chemical composition of the 3C-SiC/diamond interfaces. Finally, TDTR mapping is utilized to visualize the uniformity and quality of the SiC/diamond interfaces after annealing at 1100 °C.

## 2. Results

In this study, ~900-nm-thick 3C-SiC thin films are bonded with diamond substrates utilizing the SAB technique. More details are provided in the Methods section. Notably, a 10-nm-thick amorphous Si layer is deposited onto the diamond substrate via RF sputtering prior to the bonding process. Two bonded 3C-SiC/diamond samples undergo annealing at temperatures of 800 °C (Sample 2) and 1100 °C (Sample 3), respectively, while one sample remains bonded at room temperature (Sample 1). High-resolution transmission electron microscopy (HRTEM), accompanied by fast Fourier transform (FFT), is employed to characterize the bonded interfaces of the three samples, as depicted in Fig. 1(a). For the as-bonded sample, the interfacial region thickness between the 3C-SiC thin film and the diamond substrate measures ~10 nm. The FFT image of the as-bonded interface reveals the presence of an amorphous interlayer, characterized by the absence of long-range atomic order. Upon annealing at 800 °C and 1100 °C, significant alterations in the interfacial nanostructures of the 3C-SiC/diamond interfaces are observed. After annealing at 800 °C, the interfacial region reduces to ~9 nm, exhibiting a polycrystalline nature, as indicated by the FFT pattern. After annealing at 1100 °C, the interfacial region remains comparable to that observed after annealing at 800 °C. Crucially, the interlayer transforms into a single-crystalline state at 1100 °C, evidenced by the discernible lattice fringes associated with the 3C-SiC{111} plane and well-defined diffraction patterns in the FFT images.

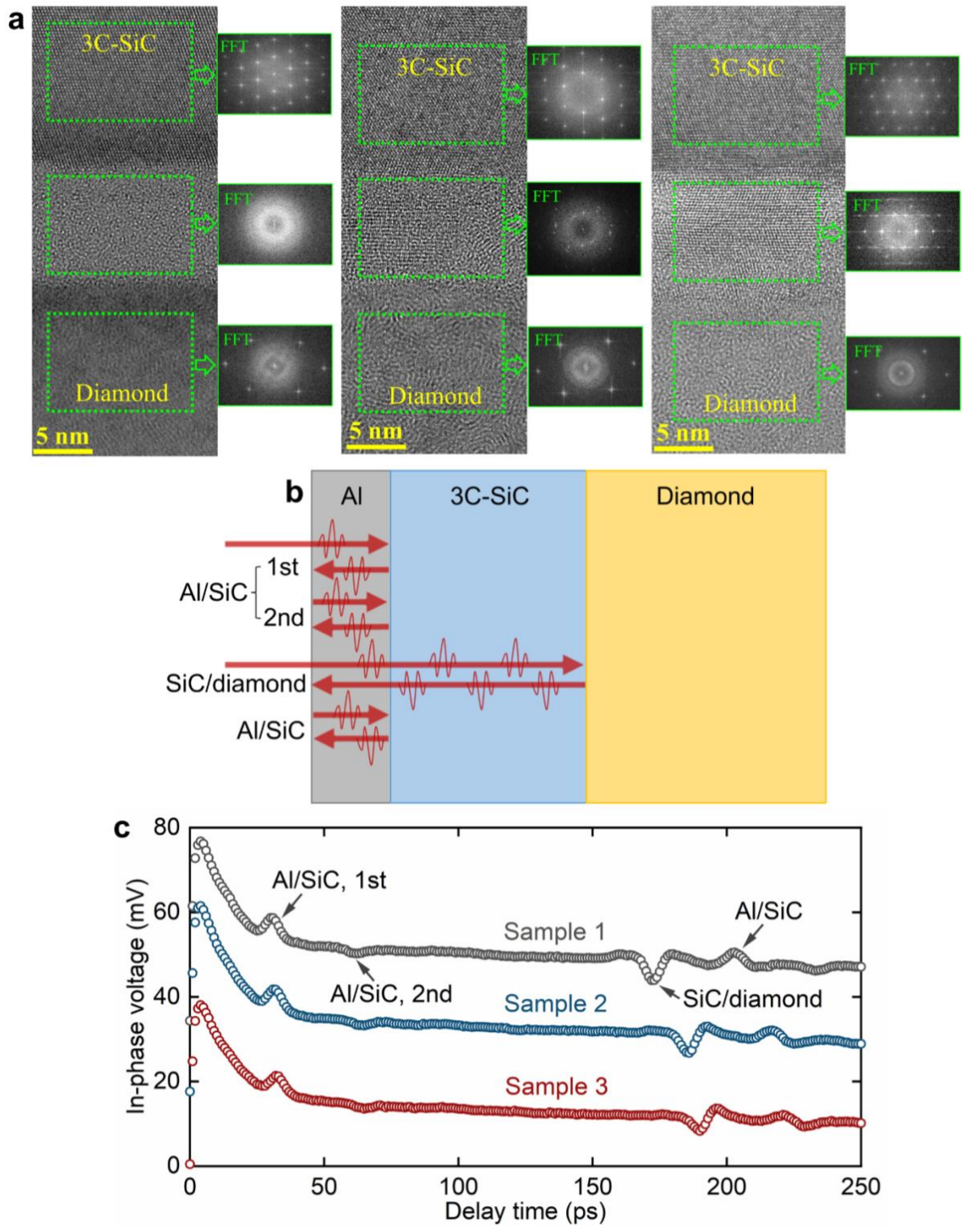

**Figure 1. Interfacial structure variations of 3C-SiC/diamond samples before and after high-temperature annealing. a** The HRTEM images and FFT images of the as-bonded, 800 °C-

annealed, and 1100 °C-annealed interfaces. The scale bars are 5 nm. The interlayer becomes single-crystalline after annealing at 1100 °C due to the observation of clear 3C-SiC{111} lattice fringes and diffraction patterns, which indicates a chemical reaction between interfacial Si and diamond in high-temperature annealing. **b** Schematic diagram of acoustic wave reflection in Al/3C-SiC/diamond multilayer structure. **c** Picosecond ultrasonics measurements on the bonded 3C-SiC/diamond samples. The shape of the echoes shows the acoustic impedance mismatch of materials near the interfaces. The transformation in the shape of the SiC/diamond echo from the as-bonded to the annealed state is due to the formation of the SiC layer.

The HRTEM results reveal an interfacial chemical reaction occurring during high-temperature annealing between the interfacial Si atoms and C atoms from the diamond interface. Consequently, the interlayer undergoes a transformation into a crystalline 3C-SiC structure at 1100 °C. Additional materials characterizations of the interfacial structures are provided in the Supplementary Information (SI). Remarkably, excellent bonding between the 3C-SiC thin film and the diamond substrate persists even after high-temperature annealing up to 1100 °C, attributed to the close thermal expansion coefficients of 3C-SiC and diamond from room temperature to 1100 °C.[38,39]

The as-bonded and annealed 3C-SiC/diamond interfaces are further characterized using picosecond ultrasonics. These measurements offer insights into the acoustic wave reflection within the multilayered sample structure.[40,41] Figure 1b depicts the acoustic wave reflection within the Al/3C-SiC/monocrystalline-diamond structure (with the Al thin film acting as the transducer), while Fig. 1c presents the corresponding picosecond acoustic signals. Acoustic waves transmit and reflect at each interface, allowing for the determination of layer thicknesses based on the travel time and sound velocity of the respective materials. Across all three samples, echoes observed at approximately 30, 60, and 210 ps correspond to acoustic wave reflections at the Al/SiC interfaces, while echoes around ~180 ps for all samples stem from reflections at the SiC/diamond interfaces. The thicknesses of the three 3C-SiC thin films are determined as 877 nm (Sample 1), 956 nm (Sample 2), and 984 nm (Sample 3), respectively.

The shape of the echo is impacted by both the morphology of the interface and the acoustic impedances of the constituent materials. Acoustic impedance (Z) is defined as $\rho v$, where $\rho$ represents density and v denotes sound velocity.[42] Consequently, the evolution in the shape of the SiC/diamond echoes from as-bonded to annealed indicates structural variation due to the interfacial reaction during annealing. A valley-shaped acoustic echo indicates acoustic wave reflection at interfaces where acoustic waves travel from high acoustic impedance to low acoustic impedance, while a peak-shaped echo indicates reflection at interfaces where acoustic waves travel from low acoustic impedance to high acoustic impedance. The acoustic echoes in Fig. 1c suggest a mixture of valley-shaped and peak-shaped echoes, corresponding to the SiC/interfacial layer interface and the interfacial layer/diamond interface. High-temperature annealing initiates a chemical reaction between Si and diamond, leading to the formation of defective crystalline SiC within the interfacial region. As the annealing temperature increases, the amplitude of the valley decreases while the amplitude of the peak increases, confirming the formation of the high acoustic impedance SiC layer and the disappearance of the low acoustic impedance Si layer. Moreover, notable wave reflection on the SiC/interfacial-SiC interface suggests the interfacial SiC is Si-rich and has lower quality than bulk SiC.

The thermal boundary conductivities (TBCs) of the 3C-SiC/monocrystalline diamond interfaces are measured using femtosecond laser-based time-domain thermoreflectance (TDTR), as illustrated in Fig. 2.[43] A modulated pump beam periodically heats the sample surface while a delayed probe beam detects temperature variation via thermoreflectance, as shown in Fig. 2a. The higher the TDTR ratio is, the higher the TBC is (Fig. 2b). More details about TDTR measurements are provided in the Methods section. The TBCs for the as-bonded interface, as well as those annealed at 800 °C and 1100 °C, are determined to be 34, 55, and 111 MW/m$^2$-K, respectively, reflecting improvements of 62% and 226% after annealing at 800 °C and 1100 °C, respectively. Additionally, TBC measurements are performed on three other 3C-SiC/polycrystalline diamond samples, revealing average TBC values of 38, 92, and 150 MW/m$^2$-K for the as-bonded samples and those annealed at 800 °C and 1100 °C, respectively. The TBCs of the SiC/polycrystalline-diamond interfaces exhibit improvements of 142% and 295% after annealing at 800 °C and 1100 °C, respectively. The TBCs of the 3C-SiC/polycrystalline diamond interfaces show much larger spot-spot variation than the 3C-SiC/monocrystalline diamond interfaces, as shown in Fig. 2c. We attribute this variation to the effect of the crystalline orientations of diamond grains on bonding. We also observe the average TBC of the 3C-SiC/polycrystalline diamond interfaces is higher than those of the 3C-SiC/monocrystalline diamond interfaces.

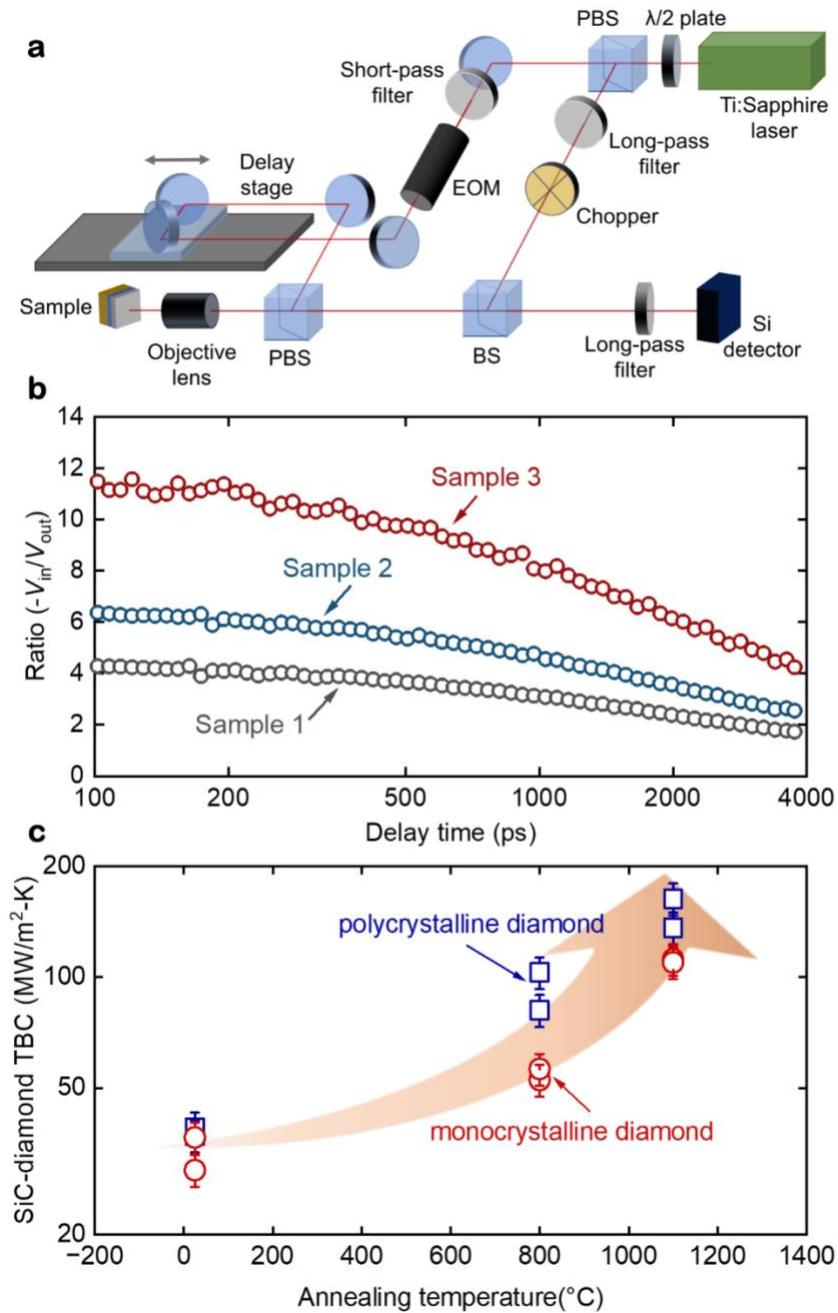

**Figure 2. Thermal measurements of 3C-SiC/diamond samples. a** Schematic diagram of TDTR setup. The pump beam, modulated at a fixed frequency, heats the sample to induce a temperature rise on the sample surface, while the probe beam is reflected from the sample surface after a delay time, carrying thermoreflectance signals. **b** TDTR ratio signals of the 3C-SiC/monocrystalline-

diamond samples. **c** TBCs of 3C-SiC/diamond interfaces as a function of the annealing temperature. An error bar of 10% is included.

Figure 3a presents the measured thermal conductivity of 3C-SiC thin films, juxtaposed with other semiconductor films integrable near room temperature. Room-temperature heterogeneous integration of high thermal conductivity materials becomes imperative since certain components in electronic devices such as metal contacts cannot endure high temperatures. The thermal conductivity of the 3C-SiC films in our study is among the highest for all semiconductor films integrable near room temperature with similar thickness. Additionally, Fig. 3b illustrates a comparison of the measured 3C-SiC/diamond TBC with those of other bonded diamond interfaces. The TBC of 3C-SiC/diamond interfaces sets a new record among all bonded diamond interfaces, facilitating the full utilization of diamond's ultrahigh thermal conductivity for electronics cooling.

Analysis of the phonon density of states (DOS), as depicted in Fig. 3c, indicates that phonon scatterings due to the presence of the Si interlayer before annealing likely pose significant limitations on the SiC/diamond TBC, attributed to poor DOS overlap between Si and diamond. The annealing process at 1100 °C leads to the formation of interfacial crystalline 3C-SiC, resulting in improved TBC partly owing to enhanced DOS overlap between 3C-SiC and diamond. Additionally, crystalline SiC possesses higher thermal conductivity than amorphous Si at the interfacial region, contributing to enhanced TBC. As illustrated in Fig. 3d, the TBC enhancement of 3C-SiC/diamond interfaces after annealing at 1100 °C sets a new record (250% – 300%) among all grown and bonded interfaces.[27,30,44–47] The high-temperature stability and elevated TBC are favorable for further high-temperature growth or fabrication of devices on these thermally conductive composite substrates.

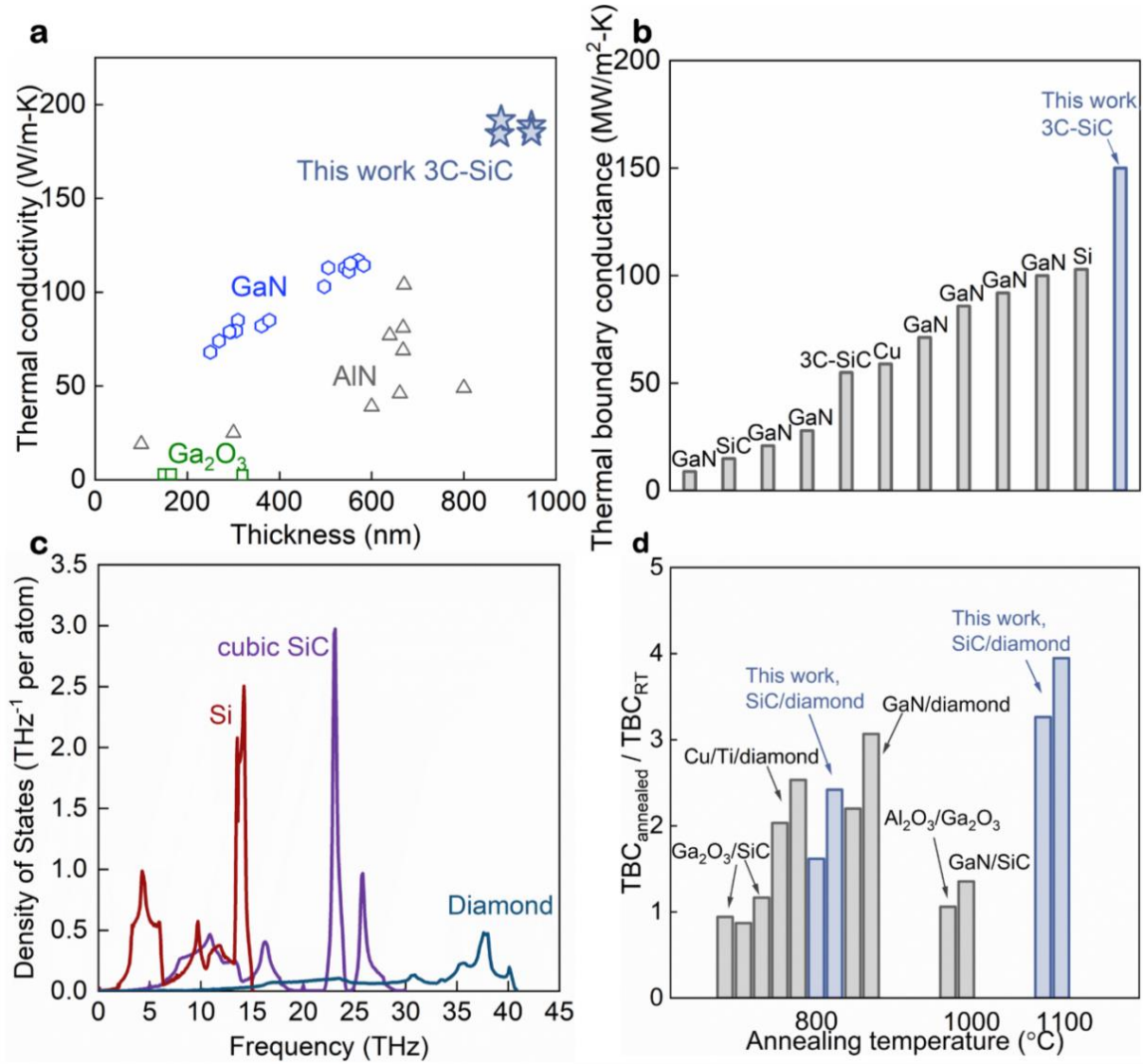

**Figure 3. Thermal properties of 3C-SiC/diamond heterostructures. a** Thermal conductivity of 3C-SiC thin films, compared with those of other semiconductor films which can be integrated near room temperature.[27,47–49]. **b** TBCs of bonded diamond interfaces.[28,30,34,36,37,50–52] **c** Phonon density of states (DOS) of Si, cubic SiC, and diamond.[53–55] The transition from Si to SiC during annealing allows for increased SiC/diamond TBC. **d** TBC enhancement after annealing.[27,30,44–47]

Integration of semiconductors with diamond often suffers from interface non-uniformity or even unbonded areas. Non-uniform interfaces, stemming from bonding non-uniformity or subsequent high-temperature processes, can significantly impede mass production yield.[56] To further explore the uniformity of SiC/monocrystalline-diamond interfaces annealed at 1100 °C, we apply the TDTR mapping technique. Figure 4a presents a schematic diagram of the 2-axis high-resolution positioning stage utilized in the study. To select the fixed delay time during TDTR mapping, we depict the sensitivities of Al/SiC TBC, 3C-SiC thermal conductivity, and SiC/diamond TBC as a function of delay time in Fig. 4b. Notably, the sensitivity of Al/SiC TBC approaches zero at short delay times but gradually increases at long delay times. Consequently, in TDTR mapping, the delay time is set at 230 ps, corresponding to the point where the sensitivity of Al/SiC TBC is negligible. Moreover, at a delay time of 230 ps, the sensitivity of SiC/diamond TBC becomes predominant in the TDTR ratio signal. This indicates that variations in TDTR ratio signals across the scanned area primarily stem from changes in the TBC of the SiC/diamond interface.

The distribution of TBC across the scanned area of 600 μm × 600 μm for the 3C-SiC/diamond interface is presented in Fig. 4c. Analysis reveals that 96% of TBC values fall within the range of 95 – 115 MW/m$^2$-K, confirming the presence of a uniform and robust bonding interface between the SiC thin film and the diamond substrate, leading to consistently high TBC values over a large area. Any irregular points displaying low TBCs are attributed to relatively weak bonding at the interface, typically occurring within a 20 μm × 20 μm area. Notably, very few relatively weak bonding areas are observed within this region. Additionally, Fig. 4d illustrates the surface reflectivity measured by the DC detector voltage. It is noteworthy that the reflectivity of the sample surface consistently exceeds 96% across the scanned area, indicating that points with poor TBC are primarily due to weak bonding within the scanned SiC/diamond interface rather than surface contaminations. We expect the bonded 3C-SiC/polycrystalline diamond interfaces are supposed to less uniform than the 3C-SiC/monocrystalline diamond interfaces.

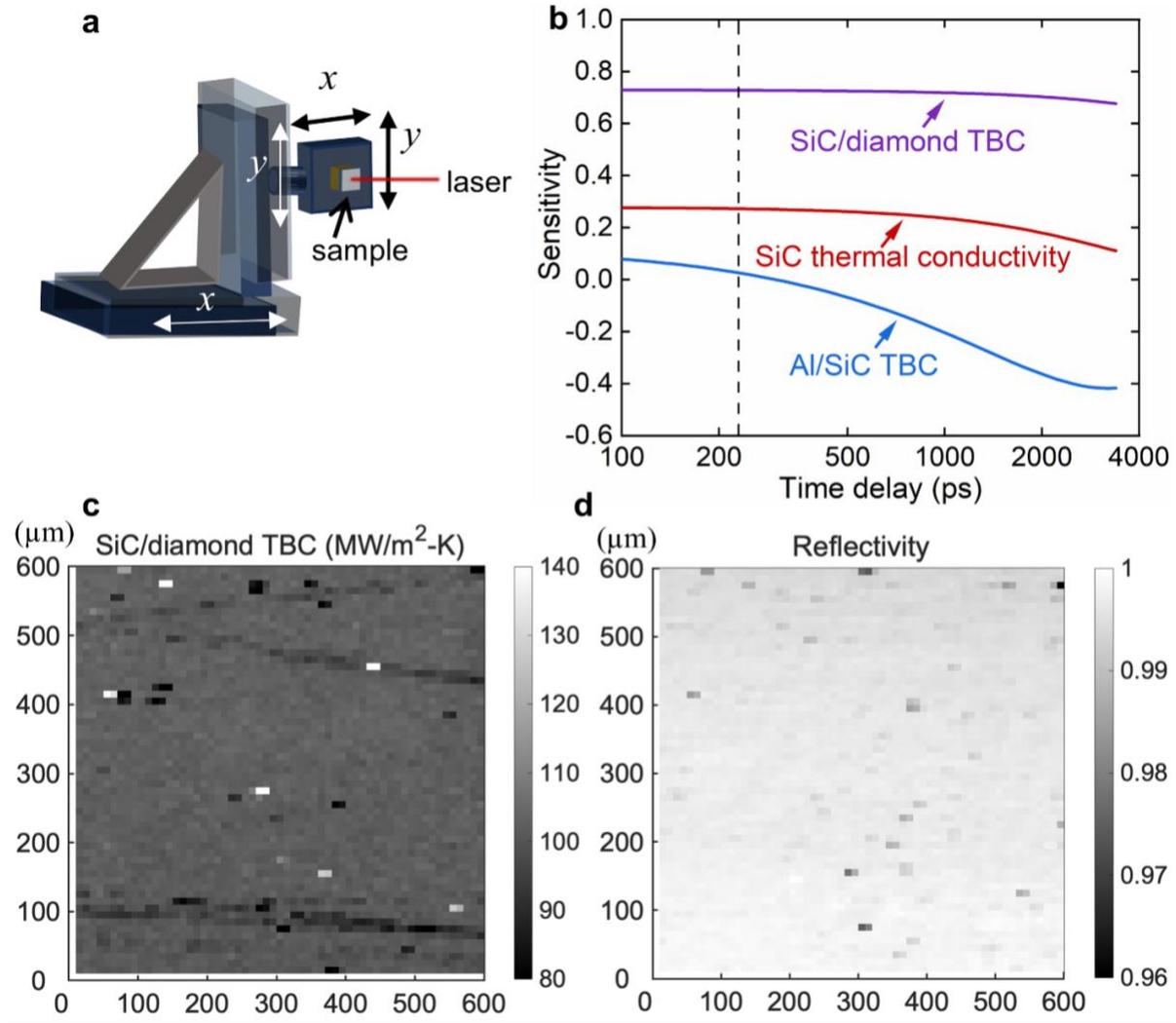

**Figure 4. TBC visualization on the 3C-SiC/diamond interface annealed at 1100 °C. a** Schematic diagram of TDTR mapping measurements. The 2-axis sample stage drives the sample holder to move row by row. **b** Sensitivity of TDTR measurements on the SiC/diamond sample. The delay time is fixed at 230 ps in TDTR mapping because the sensitivity of the Al/SiC TBC is zero at this delay time. **c** Mapped TBCs in the scanned area. **d** Distribution of detector DC voltages in the scanned area. The scanned area is 600 μm×600 μm.

The TBC visualization results confirm the presence of uniform bonding interfaces with high TBCs between the 3C-SiC thin film and the diamond substrate, capable of withstanding high temperatures up to 1100°C. These findings hold promise for device growth or fabrications at high temperatures, enabling the integration of devices and high thermal conductivity substrates with thermally conductive interfaces. This sheds light on a comprehensive solution to address thermal management challenges in next-generation electronics such as chiplets, power, and RF electronics.

## 3. Conclusion

This study shows heterogeneous integrated semiconductor heterostructures: 3C-SiC and mono/polycrystalline diamond substrates. The 3C-SiC thin films exhibit record-high thermal conductivity among all semiconductor films integrable near room temperature with comparable thicknesses. Notably, remarkable enhancements in TBC of the 3C-SiC/diamond interfaces are observed after high-temperature annealing, which is record-high among all grown and bonded semiconductor interfaces. The 3C-SiC-diamond TBC after annealing is also record-high among all bonded diamond interfaces. It is attributed to that, after annealing at 1100°C, the amorphous silicon interlayer undergoes transformation into crystalline SiC through a chemical reaction between interfacial Si and C atoms from the diamond substrates. Comparatively, the TBC of the SiC/monocrystalline-diamond interface improves by 226% after high-temperature annealing, while that of SiC/polycrystalline-diamond increases by 295%. Picosecond ultrasonics measurements validate structural changes after annealing at different temperatures. Furthermore, the TBCs revealed by TDTR mapping demonstrate high uniformity, with 96% of values falling within the range of 95 to 115 MW/m$^2$-K. The stability of interfacial bonding and high TBC offer a promising pathway for device fabrication on these composite substrates, promising excellent cooling performance.

## 4. Experimental methods

### 4.1. Integrated 3C-SiC/diamond heterostructures by SAB

To fabricate the samples in this study, 3C-SiC epitaxial layers are initially grown on Si substrates, followed by transfer to supporting substrates after removal of the raw Si substrates using fluoric acid etching. Subsequently, both the 3C-SiC films and the diamond substrates are polished to achieve a root mean square (RMS) roughness of less than 1 nm. The 3C-SiC surface undergoes cleaning with acetone and isopropyl alcohol, followed by drying with N$_2$ flow. Similarly, the diamond substrate surface is cleaned using a sulfuric acid–hydrogen peroxide mixture, rinsed with deionized water, and dried with N$_2$.[31] Ar ion beams are then employed to clean and activate the to-be-bonded surfaces, creating dangling bonds within an ultrahigh vacuum environment. The two substrates are subsequently pressed together at room temperature to form strong covalent bonds.[34,57–59] Following the bonding process, the supporting substrate is removed. More details about the bonding procedure can be found in the SI.

### 4.2. Atomic scale materials characterizations

Transmission electron microscopy (TEM) was performed with a JEOL JEM-2200FS analytical microscope at an acceleration voltage of 200 kV. The TEM samples were prepared using a focused ion beam (FIB) with a Helios NanoLab 600i Dualbeam system from Thermo Fisher Scientific, Inc. More materials characterizations can be found in SI.

### 4.3. Thermal measurements by TDTR

A 100-nm Al thin film is deposited on the sample surface to serve as the transducer layer in TDTR measurements. The TDTR setup divides the mode-locked Ti:Sapphire laser into a pump beam and a probe beam. The pump beam, modulated at a fixed frequency, heats the sample to induce a temperature rise on the sample surface, while the delayed probe beam is reflected from the sample surface to measure the temperature variation via thermoreflectance. Both the pump and probe beams are focused onto the sample using an objective lens, with the reflected probe beam detected by a photodiode. An RF lock-in amplifier captures the thermoreflectance signal synchronized with the pump modulation frequency. The ratio of the in-phase voltage ($V_{in}$) to the out-of-phase voltage signal ($V_{out}$, ratio = -$V_{in}$/$V_{out}$) is collected for subsequent data analysis. The TDTR ratio signals are then fitted to an analytical solution of a multilayered heat transfer model to determine the unknown thermal parameters.[60] Initially, the thermal conductivities of the bare monocrystalline/polycrystalline diamond substrates are measured. When measuring the 3C-SiC/diamond samples, a frequency of 1.9 MHz is selected to ensure a sufficient thermal penetration depth, thereby enabling sensitive measurements of the SiC/diamond TBC. The thermal penetration depth ($d$) is related to the pump modulation frequency ($f$) by $d = \sqrt{\Lambda/\pi C f}$, where $\Lambda$ is the thermal conductivity and $C$ is the volumetric heat capacity.

### 4.4. TBC visualization by TDTR mapping

For both TDTR single-point and mapping measurements, the 1/e² radius of the laser was 9.7 μm for the 5× objective and 4.9 μm for the 10× objective. In TDTR mapping, we utilized a piezoelectric two-dimensional positioning stage (Physik Instrumente, Q-545.240) with the 10× objective lens.[61] An area of 600 μm×600 μm was scanned on the 3C-SiC-diamond sample annealed at 1100 °C with a step size of 10 μm. The positioning stage was programmed to scan the sample surface row by row, pausing for 0.5 s at each new position to collect data before moving to the next position. Additionally, the waiting time was doubled at the beginning of every new row.

To mitigate the influence of the Al/SiC TBC variation, mapping signals were collected at a fixed delay time when the sensitivity of the Al/SiC TBC reached zero. The sensitivity in TDTR measurements is defined as:

$$S_\alpha = \frac{d\ln(-V_{in}/V_{out})}{d\ln\alpha},$$

where α represents the fitting parameter.[62,63] The sensitivity of the Al/SiC TBC is nearly zero at short delay times but increases at long delay times. Consequently, in TDTR mapping, the delay time was fixed at 230 ps when the sensitivity of the Al/SiC TBC is zero.

We converted the mapping of TDTR ratio signal into a mapping of the 3C-SiC/diamond TBC.[64] A relationship is established between the ratio signals and the expected TBC values, and then the experimentally scanned ratio signals are converted into corresponding TBC values. During the data analysis of the modeled ratios, the TBC of the Al/SiC interface, the thermal conductivity of the SiC, and the diamond substrate are held constant at 151 MW/m²-K, 193 W/m-K, and 1964 W/m-K, respectively. Subsequently, the relationship is calculated as follows:

$$TBC(\text{MW/m}^2\text{-K}) = 7.432e^{-6}R^3 - 1.567e^{-4}R^2 + 0.008603R - 0.01886.$$

Where R is the mapped ratio signals, and TBC is the mapped SiC/diamond TBC.

**Conflict of Interest**

The authors declare no conflict of interest.

**Acknowledgments**

X.J. and Z.C. would like to thank Prof. David Cahill for his valuable comments on the manuscript. Z.C. would like to thank the support by "The Fundamental Research Funds for the Central Universities, Peking University". Part of this work was based on results obtained from a project, JPNP20004, subsidized by the New Energy and Industrial Technology Development Organization (NEDO). The TEM samples were fabricated at The Oarai Center and the Laboratory of Alpha-Ray Emitters in IMR under the Inter-University Cooperative Research in IMR of Tohoku University (202212-IRKMA-0016).